\documentclass{iaus}
\usepackage{graphicx}
\usepackage{graphicx,bm,url}
\usepackage{natbib}

\newcommand{\EQ}{\begin{equation}}
\newcommand{\EN}{\end{equation}}
\newcommand{\EQA}{\begin{eqnarray}}
\newcommand{\ENA}{\end{eqnarray}}

\newcommand{\bra}[1]{\langle #1\rangle}

\newcommand{\mean}[1]{\overline #1}

{}
{}
{}

{}
{}
{}
{}
{}
{}
{}
{}
{}
{}
{}
{}
{}
{}
{}
{}
{}
{}
{}
{}

{}
{}
{}

{}

{}
{}

\newcommand{\pd}{\partial}
\newcommand{\chit}{\chi_{\rm t}}
\def\onehalf{{\textstyle{1\over2}}}
\def\onethird{{\textstyle{1\over3}}}

\newcommand{\DIV}{\bm{\nabla} \cdot }
\newcommand{\const}{{\rm const}  {}}

\def\Ma{\mbox{\rm Ma}}

\def\Pra{\mbox{\rm Pr}}

\def\Rey{\mbox{\rm Re}}
\def\Rem{\mbox{\rm Rm}}

\def\kef{k_{\rm f}}

\def\urms{u_{\rm rms}}

\def\etat{\eta_{\rm t}}

\def\onethird{{\textstyle{1\over3}}}

\newcommand{\yapj}[3]{ #1, {ApJ,} {#2}, #3}

\newcommand{\yapjl}[3]{ #1, {ApJL,} {#2}, #3}

\newcommand{\yan}[3]{ #1, {Astron.\ Nachr.,} {#2}, #3}

\newcommand{\yana}[3]{ #1, {A\&A,} {#2}, #3}

\newcommand{\yjetp}[3]{ #1, {Sov.\ Phys.\ JETP,} {#2}, #3}

\newcommand{\ymn}[3]{ #1, {MNRAS,} {#2}, #3}

\newcommand{\ysph}[3]{ #1, {Solar Phys.,} {#2}, #3}

\newcommand{\ypre}[3]{ #1, {Phys.\ Rev.\ E,} {#2}, #3}

\title[Flux concentrations in turbulent convection]
{Flux concentrations in turbulent convection}

\author[P.\ J.\ K\"apyl\"a, A.\ Brandenburg, N.\ Kleeorin,
M.\ J.\ Mantere \&  I.\ Rogachevskii] {Petri J.\
K\"apyl\"a$^{1,2}$, Axel Brandenburg$^{2,3}$,
Nathan Kleeorin$^{4,2}$, Maarit J.\
Mantere$^{1}$, Igor Rogachevskii$^{4,2}$ }

\affiliation{
$^1$Department of Physics, PO Box 64, FI-00014 University of Helsinki, Finland\\
$^2$Nordita, KTH Royal Institute of Technology and Stockholm University, Roslagstullsbacken 23, SE-10691 Stockholm, Sweden\\
$^3$Department of Astronomy, Stockholm University, SE-10691 Stockholm, Sweden\\
$^4$Department of Mechanical Engineering, Ben-Gurion University of the Negev, PO Box 653, Beer-Sheva 84105, Israel
}

\pubyear{2013}
\volume{294}
\pagerange{1--6}
\date{$ $Revision: 1.49 $ $}
\jname{Solar and Astrophysical Dynamos and Magnetic Activity}
\editors{A.G. Kosovichev, E.M. de Gouveia Dal Pino \& Y. Yan, eds.}

\begin{document}

\maketitle

\begin{abstract}
  We present preliminary results from high resolution
  magneto-convection simulations where we find the formation of flux
  concentrations from an initially uniform magnetic field. We compute
  the effective magnetic pressure but find that the concentrations
  appear also in places where it is positive. The structures appear in
  roughly ten convective turnover times and live close to a turbulent
  diffusion time. The time scales are compatible with the negative
  effective magnetic pressure instability (NEMPI), although structure formation
  is not restricted to regions where the effective magnetic pressure is
  negative.

\end{abstract}

\firstsection

\section{Introduction}
The current paradigm of sunspot formation is based on the idea of
buoyant rise of flux tubes from the base of the solar convection zone
to the surface of the Sun \citep{P55}. This process is parameterised
in the widely used flux transport dynamo models \citep[e.g.][]{DC99}
in the form of a non-local $\alpha$-effect where the strong (around
$10^5$G) toroidal magnetic fields in the tachocline give rise to
poloidal fields at the surface. This poloidal magnetic field is then
advected by meridional circulation back to the tachocline where it is
amplified.

These concepts face several theoretical difficulties, however,
including the storage and generation of strong magnetic fields beneath
the convection zone \citep[e.g.][]{GK11}, and the stability of the
tachocline in the presence of such strong fields
\citep{ASR05}. Furthermore, observations of sunspot rotation suggest
that they might be a shallow phenomenon possibly occurring within the
near surface shear layer \citep{B05}. This requires a new mechanism to
form sunspots.

Theoretical works have shown that suitable turbulence can have a
negative contribution to the magnetic pressure
\citep[e.g.][and references
therein]{KRR90,KMR96,KR94,RK07}. This effect leads to
the negative effective magnetic pressure instability (NEMPI) where
even uniform, sub-equipartition, magnetic fields can form flux
concentrations. This is compatible in view of the results from direct
simulations (DNS) of convection driven dynamos where diffuse magnetic
fields are generated in all of the convection zone
\citep[e.g.][]{GCS10,KMB12}.

Recently, a lot of effort has been devoted to study this effect using
mean-field models and DNS of forced turbulence
\citep[e.g.][]{BKR10,BKKR12,KBKR12}, culminating in the
detection of NEMPI in DNS \citep{BKKMR11}.
A negative turbulent contribution to the effective (mean-field)
magnetic pressure has also been found for convection
\citep{KBKMR12} but no NEMPI has been detected so
far. Here we present results from new high
resolution convection simulations designed to be
better suited for the detection of NEMPI.

\section{Model}

We solve the compressible hydromagnetics equations,
\begin{equation}
\frac{\pd \bm A}{\pd t} = {\bm u}\times{\bm B} - \eta \mu_0 {\bm J},
\end{equation}
\begin{equation}
\frac{D \ln \rho}{Dt} = -\bm\nabla\cdot\bm{u},
\end{equation}
\begin{equation}
\frac{D\bm{u}}{Dt} = \bm{g} + \frac{1}{\rho}
\left(\bm\nabla \cdot 2\nu\rho\bm{\mathsf{S}}-\bm\nabla p + {\bm J} \times {\bm B}\right),
\end{equation}
\begin{equation}
T\frac{D s}{Dt} = \frac{1}{\rho}\left[\bm\nabla \cdot (K \bm\nabla T + \chit \rho T \bm\nabla s) + \mu_0\eta {\bm J}^2\right] + 2\nu \bm{\mathsf{S}}^2 ,
\label{equ:ss}
\end{equation}
where ${\bm A}$ is the magnetic vector potential, $\bm{u}$ is the
velocity, ${\bm B} =\bm\nabla\times{\bm A}$ is the magnetic field,
${\bm J} =\mu_0^{-1}\bm\nabla\times{\bm B}$ is the current density,
$\eta$ is the magnetic diffusivity, $\mu_0$ is the vacuum
permeability, $D/Dt = \pd/\pd t + \bm{u} \cdot \bm\nabla$ is the
advective time derivative, ${\bm g}=-g\hat{\bm e}_z=\const$ is
gravity, $\nu$ is the kinematic viscosity, $K$ is
the radiative heat conductivity, $\chi_{\rm t}$ is the unresolved
turbulent heat conductivity, $\rho$ is the density,
$s$ is the specific entropy, $T$ is the temperature, and $p$ is the
pressure. The fluid obeys the ideal gas law with $p=(\gamma-1)\rho e$,
where $\gamma=c_{\rm P}/c_{\rm V}=5/3$ is the ratio of specific heats
at constant pressure and volume, respectively, and $e=c_{\rm V} T$ is
the internal energy. The traceless rate of strain tensor
$\mbox{\boldmath ${\sf S}$}$ is given by
\begin{equation}
{\sf S}_{ij} = \onehalf (U_{i,j}+U_{j,i}) - \onethird \delta_{ij} \DIV \bm{U}.
\end{equation}

We omit stably stratified layers above and below the convection
zone. The depth of the layer is $L_z=d$ whereas the horizontal
extents are $L_x/d=10$ and $L_y/d=5$.
The boundary conditions for the flow are impenetrable and stress free,
and perfectly conducting for the magnetic field. The energy flux at
the lower boundary is fixed, and we use a black body boundary
condition given by
\begin{equation}
\sigma T^4=- c_{\rm P} \rho \chi \frac{dT}{dz} - \rho T \chit \frac{ds}{dz},
\end{equation}
where $\sigma$ is the Stefan--Boltzmann constant at the surface
\citep[cf.][]{KMB11}.

\begin{figure}[t!]\begin{center}
\includegraphics[width=\textwidth]{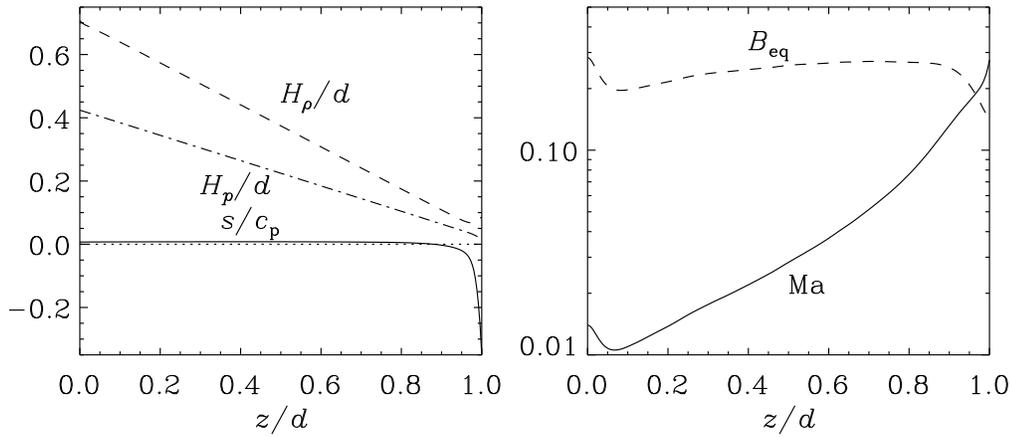}
\end{center}\caption[]{Left panel: specific entropy (solid line), and
  density (dashed) and pressure (dot-dashed) scale heights as
  functions of depth. Right
  panel: Mach number (solid line) and equipartition field strength
  from the thermally saturated regime as functions of
  depth.}\label{pstrat}\end{figure}

We use a constant $\chi=K/(c_{\rm P}\rho)$ whereas $\chit$ is zero
below $z/d<0.5$, $\chit/\chi=10$ in the range $0.5<z/d<0.9$, and
$\chit/\chi=50$ above $z/d>0.9$. The Prandtl number $\Pra=\nu/\chi$ is
equal to 10. In this setup convection transports the majority of the
flux whereas radiative diffusion is only important near the bottom of
the domain. We start a hydrodynamic progenitor run from an isentropic
stratification with density stratification of $80$. The density and
pressure scale heights, mean entropy profile, equipartition magnetic
field $B_{\rm eq}=\langle\mu_0 \rho {\bm u}^2\rangle^{1/2}$, and the
Mach number, $\Ma=\urms/c_{\rm s}$, in the thermally saturated state
of the simulation are shown in Fig.~\ref{pstrat}. In the simulation
considered here the fluid and magnetic Reynolds numbers are
$\Rey=\urms/(\nu k_1)\approx94$ and $\Rem=\urms/(\eta k_1)\approx5$,
respectively, where $\urms$ is the rms value of the volume averaged
velocity and $k_1=2\pi/d$. We use a grid resolution of
$1024\times512\times256$. The computations were performed with the
{\sc Pencil Code}\footnote{http://pencil-code.googlecode.com/}.

\begin{figure}[t!]\begin{center}
\includegraphics[width=0.49\textwidth]{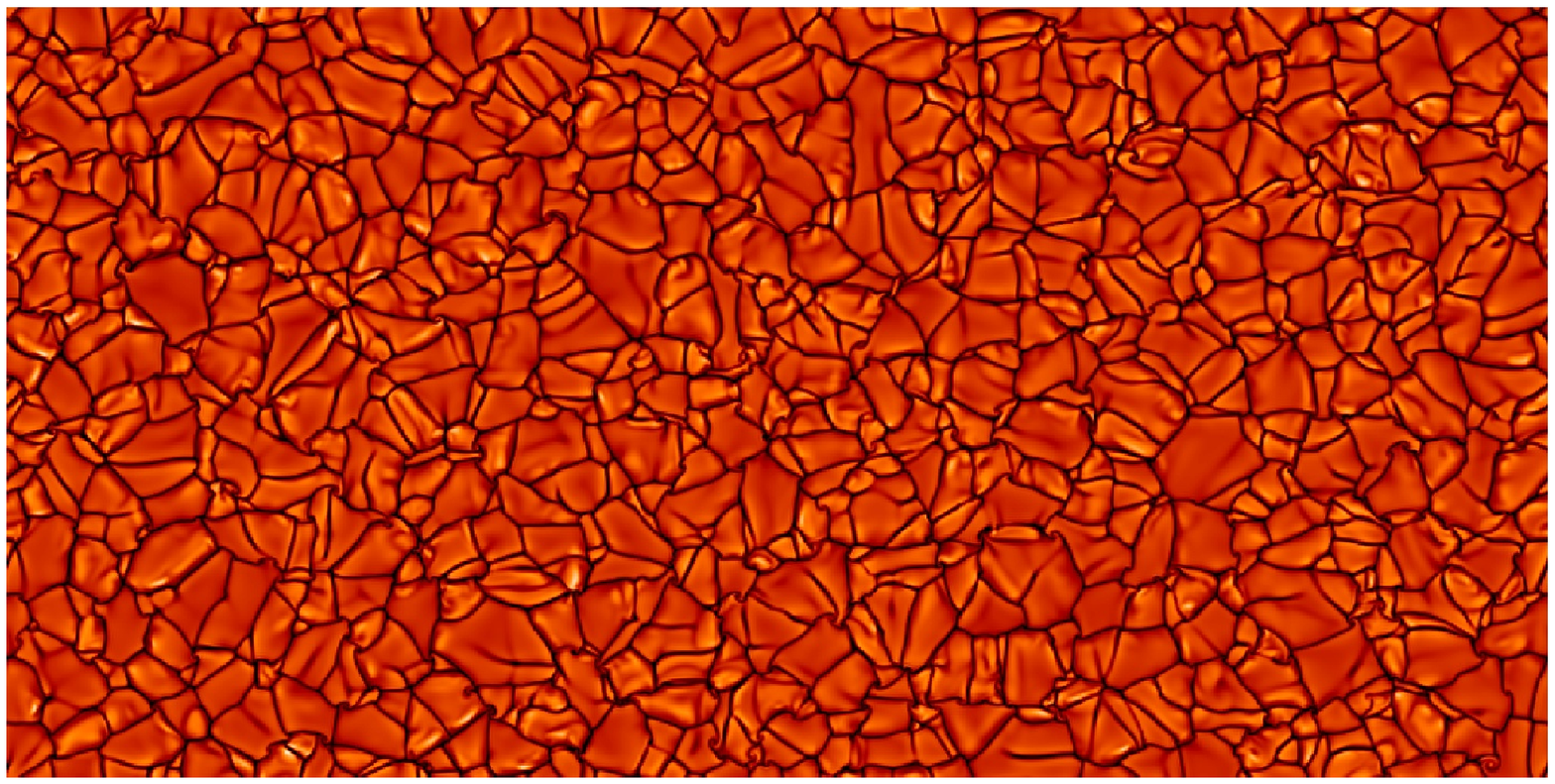}\hspace{0.1cm}\includegraphics[width=0.49\textwidth]{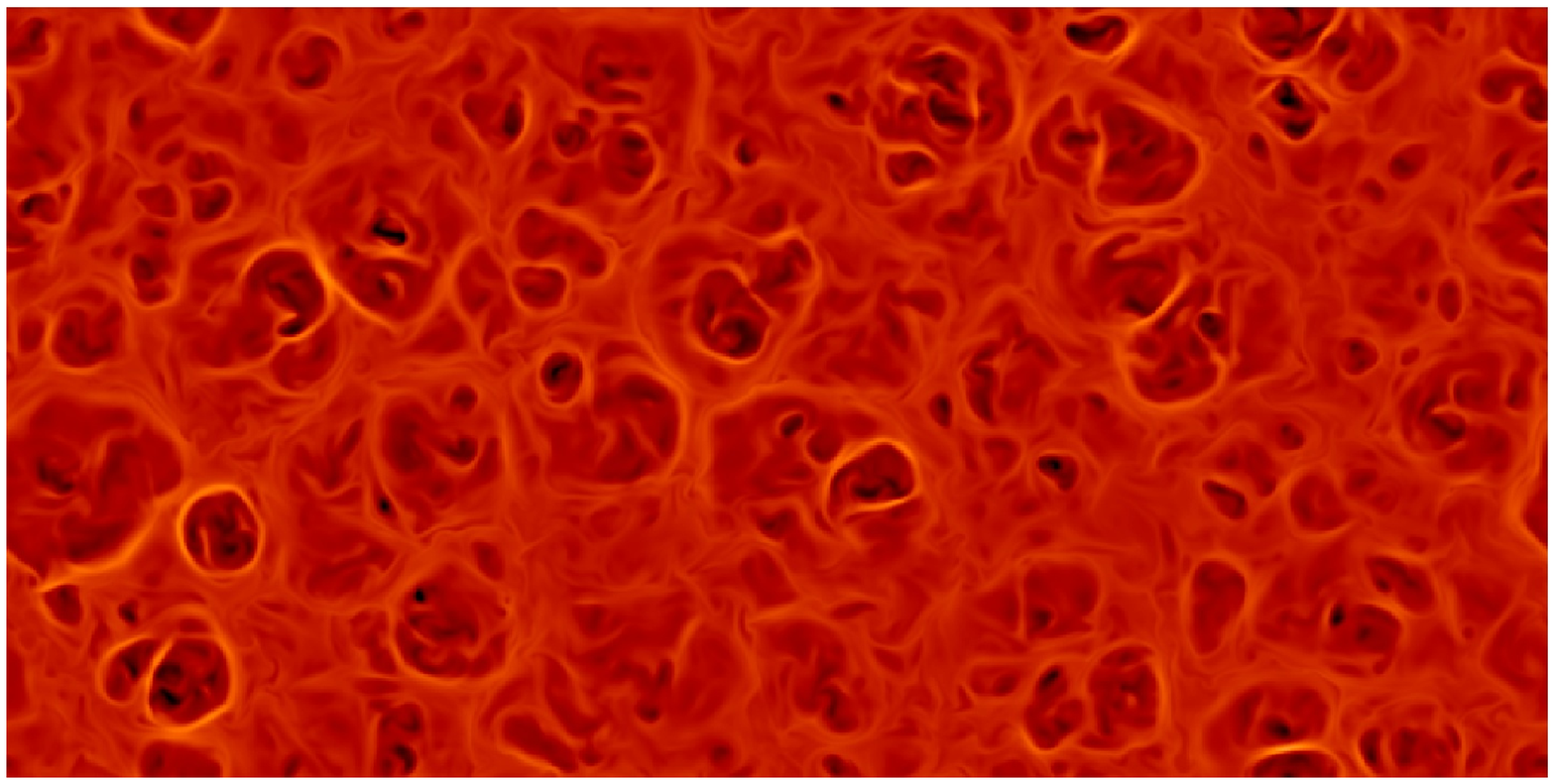}
\end{center}\caption[]{Vertical velocity $U_z$ near the surface (left
  panel) and near the bottom (right) of the domain from a
  hydrodynamical simulation. The contour levels correspond to $\pm
  0.37 \Ma$ (left) and $\pm 0.03 \Ma$
  (right).}\label{m1024x512x256a1_Uz}\end{figure}

\section{Results}

We first allow a hydrodynamic progenitor simulation to saturate after
which we impose a uniform horizontal field $B_0\hat{\bm e}_y$ with
$B_0/B_{\rm eq}\approx0.3$.
The vertical velocity $U_z$ near the surface ($z/d=0.98$) and near the
bottom ($z/d=0.02$) of the domain in the hydrodynamic run are shown in
Fig.~\ref{m1024x512x256a1_Uz}.
In forced turbulence simulations NEMPI appears when the scale
separation between the forcing scale and the box size is of the order
of 15 \citep{BKKMR11}. In our convection setup this is probably
satisfied near the surface but not near the bottom of the domain.

\begin{figure}[t!]\begin{center}
\includegraphics[width=0.97\textwidth]{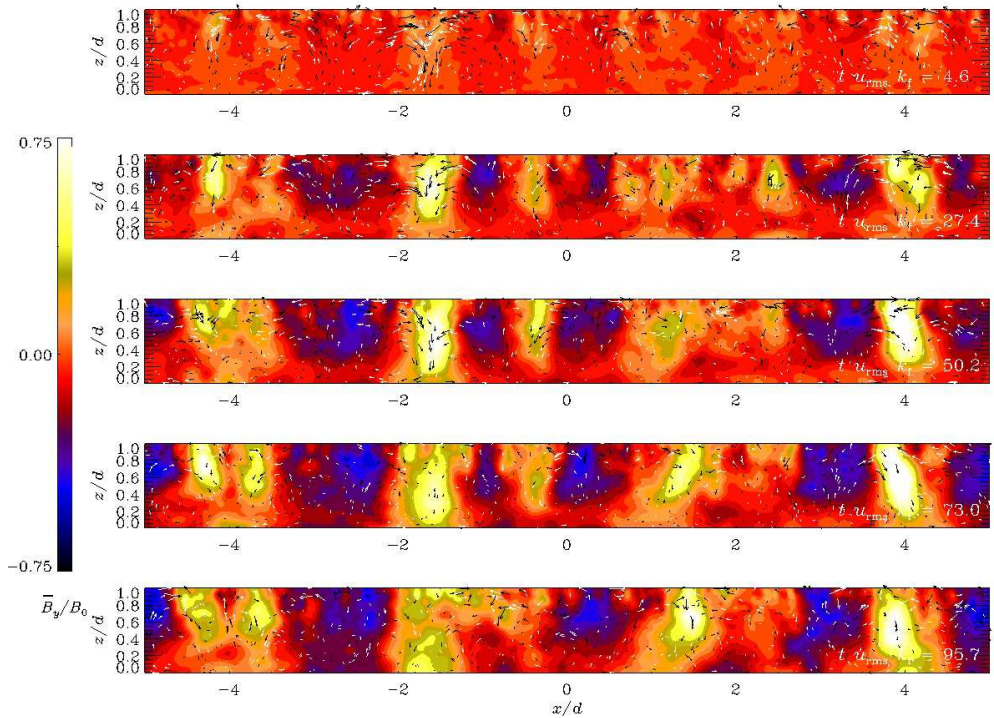}
\end{center}\caption[]{Mean magnetic field component $\mean{B}_y$ from
  five times separated by
  $23\ t \urms \kef$
  after initializing the
  magnetic field. The black and white arrows show the $y$-averaged
  velocity in the $(x,z)$-plane.}
\label{pby_xz}\end{figure}

\begin{figure}[t!]\begin{center}
\includegraphics[width=\textwidth]{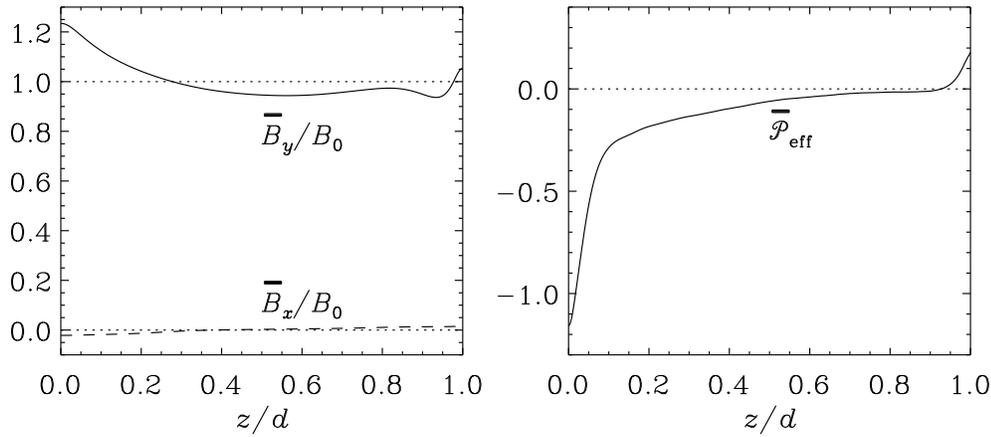}
\end{center}\caption[]{Left panel: horizontally averaged horizontal
  magnetic fields $\mean{B}_x$ and $\mean{B}_y$ in units of the
  imposed field $B_0$. Right panel: horizontal average of the
  effective magnetic pressure $\mean{\mathcal{P}}_{\rm eff}$. The
  average is taken over a period where the rms magnetic field is
  saturated.}
\label{pbyz}\end{figure}

\begin{figure}[t!]\begin{center}
\includegraphics[width=0.49\textwidth]{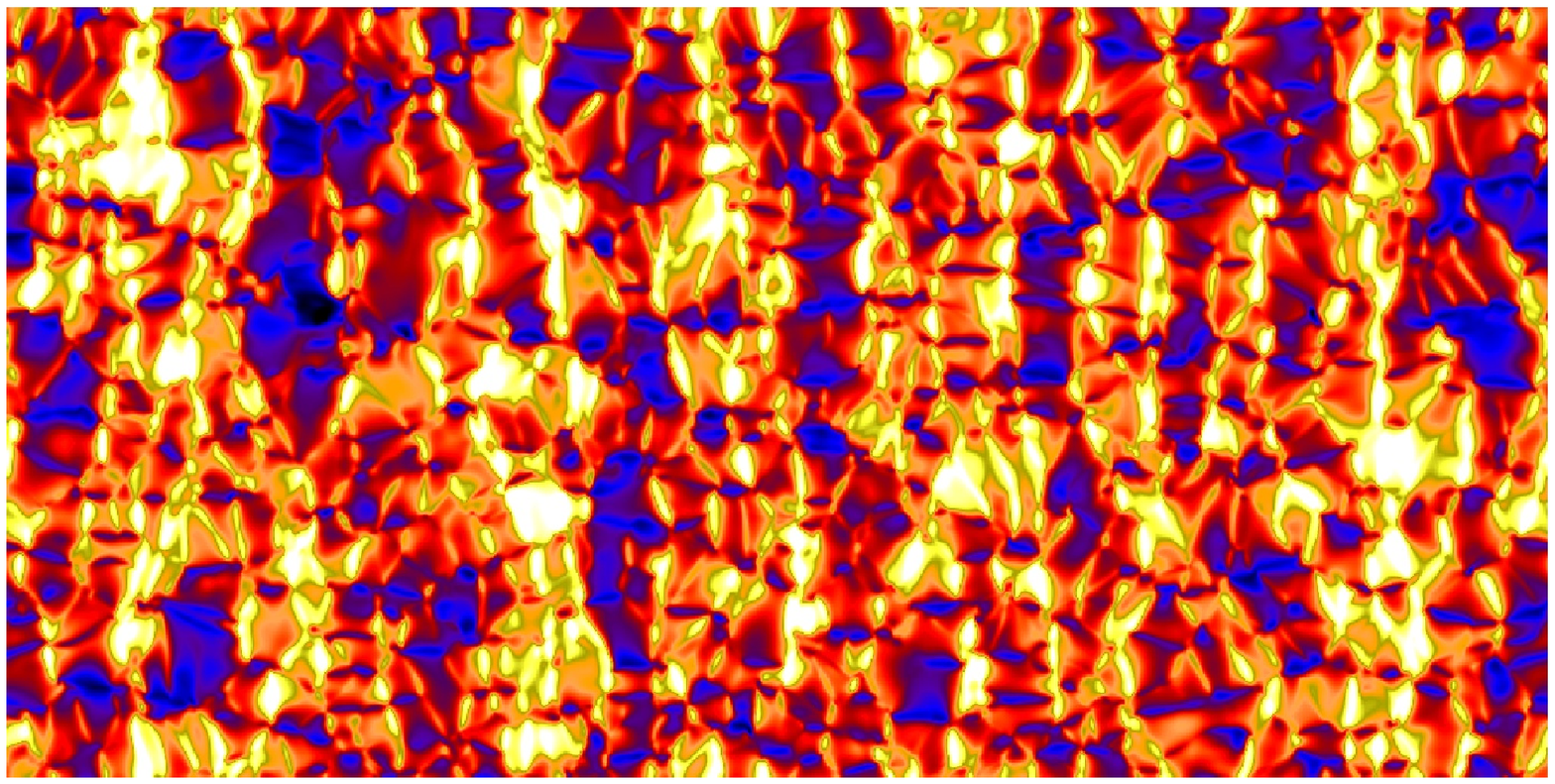}\hspace{0.1cm}
\includegraphics[width=0.49\textwidth]{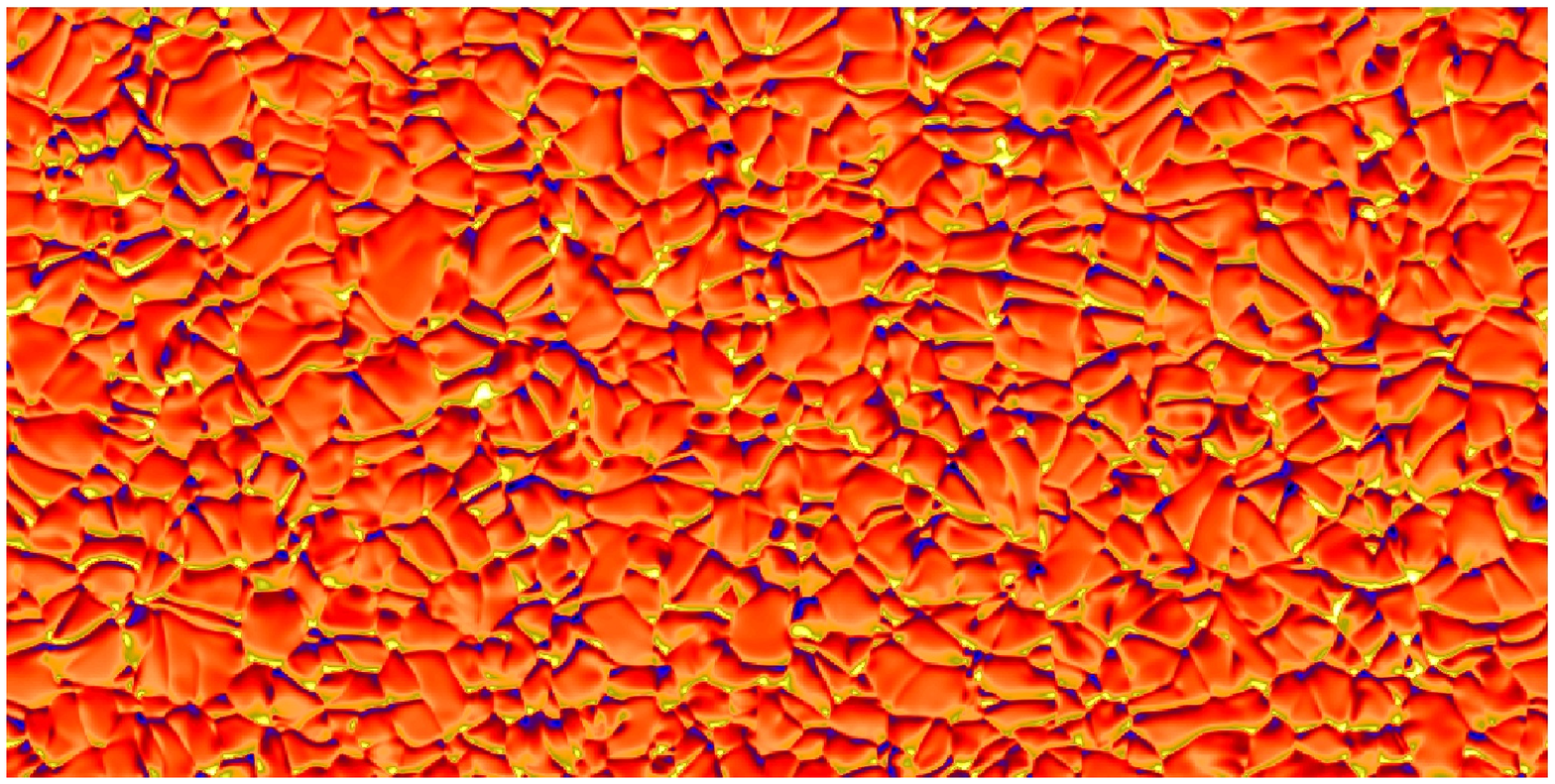}
\end{center}\caption[]{Magnetic field components $B_y$ (left panel)
  and $B_z$ (right) near the surface ($z/d=0.98$) from the same
  time as in the bottom panel of Fig.~\ref{pby_xz}. The contour levels
  correspond to $-B_0<B_y-B_0< B_0$ (left) and $-B_0<B_z< B_0$
  (right).}\label{m1024x512x256a3_By}\end{figure}

We find that large-scale structures form within ten convective
turnover times;
see Fig.~\ref{pby_xz} where
$\mean{B}_y\equiv\mean{B}_y(x,z)-\bra{B_y}_{xy}(z)$
is shown. The subscripts refer to averages over
either only $y$ or over both horizontal
directions, respectively. We remove the
horizontal mean value (see the left panel of
Fig~\ref{pbyz}) in order to make the horizontal
variation of $\mean{B}_y(x,z)$ visible. The
magnetic structures appear near the surface and
sink on a
timescale of a few tens of turnover times $\tau=(\urms \kef)^{-1}$,
where we
estimate $\kef$ by $k_\omega=\omega_{\rm rms}/\urms$, and
$\bm\omega=\bm\nabla\times{\bm u}$. Now the turbulent diffusion time
$\tau_{\rm diff}=(\etat k_1^2)^{-1}$ is roughly 180 turnover times if
we assume
$\etat=\urms/(3\kef)$ for the turbulent diffusivity. The maximum field
strength in the
concentrations is of the order of the imposed field.
The elongated magnetic structures are also weakly discernible in the
instantaneous magnetic field $B_y$, see the left panel of
Fig.~\ref{m1024x512x256a3_By}. In the vertical field $B_z$ from the
same instant, however, it is not possible to distinguish the same
structures, right panel of Fig.~\ref{m1024x512x256a3_By}. This can due
to the perfect conductor boundary condition which imposes $B_z=0$ a
the boundary.

We define the effective magnetic pressure as $\overline{\mathcal{P}}_{\rm
  eff}=\onehalf(1-q_p)\frac{\mean{B}^2}{B_{\rm eq}^2}$, where
$q_p=-2\Delta \overline{\rho u_y^2} -\Delta \overline{\bm b^2} +
2\Delta \overline{b_y^2}$, and where $\Delta$ refers to the difference
between runs with and without an imposed field. In the present case a
small-scale dynamo is absent and thus the magnetic correlations come
only from the run with the imposed field.
We find that the effective magnetic pressure $\overline{\mathcal{P}}_{\rm eff}$
is positive near the upper boundary and negative below $z/d<0.1$ with
increasingly negative values towards the bottom, see the right panel
of Fig.~\ref{pbyz}. The maxima of
$\overline{\mathcal{P}}_{\rm eff}$ are associated with maxima of
$\mean{B}_y$, whereas the minima of $\overline{\mathcal{P}}_{\rm eff}$
coincide with lower
average values of $\mean{B}_y$, see Fig.~\ref{pby_xz}.

\begin{figure}[t!]\begin{center}
\includegraphics[width=0.97\textwidth]{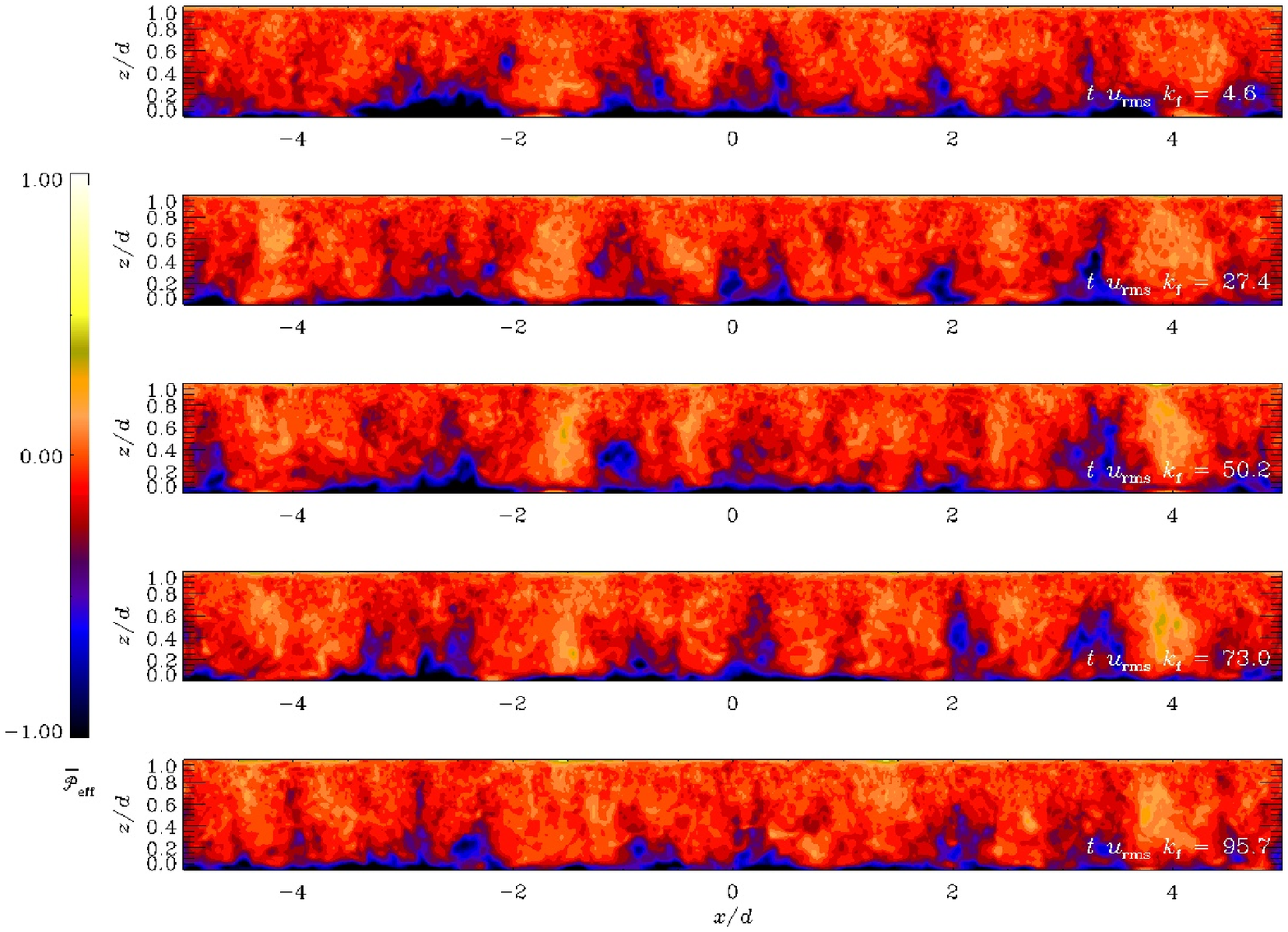}
\end{center}\caption[]{Effective magnetic pressure 
$\overline{\mathcal{P}}_{\rm eff}$ from
  the same times as in Fig.~\ref{pby_xz}.}
\label{ppeff}\end{figure}

\section{Conclusions}

Our results have shown that convection can lead
to magnetic flux concentrations by
a  mechanism that may be related to NEMPI.
A possible alternative mechanism is magnetic flux
expulsion that has previously been found to be
responsible for a segregation of magnetised and
unmagnetised regions in large aspect ratio
convection simulations with an imposed {\em
vertical} magnetic field \citep{Tao}. Our results
appear to be similar, except that here we have an
imposed {\em horizontal} magnetic field. In that
respect, it is useful to mention recent
simulations of \cite{SN12}, who inject a 1000\,G
horizontal magnetic field at the bottom of their
simulated convection domain and find after some
time the emergence of a bipolar magnetic field at
the surface. In their case the reason for the
formation of flux concentrations is argued to be
the downdrafts of the deeper supergranulation
pattern, which tend to keep the magnetic field
concentrated into flux bundles at the bottom of
their open domain. Our present simulations do not
capture this effect, because they are probably
not deep enough and our domain is impenetrative
and perfectly conducting at the bottom, excluding
therefore their mechanism as a possible
explanation. Of course, another important
difference between our simulations and those of
\cite{SN12} is the presence of a radiating
surface in their case. This might enhance
magnetic flux concentrations formed through local
suppression of convective energy flux by magnetic
fields \citep{KM00}. This might well be an
important effect that needs to be studied more
thoroughly.

\acknowledgments

The computations were performed on the facilities
hosted by the CSC -- IT Center for Science in
Espoo, Finland, which are financed by the Finnish
ministry of education. We acknowledge financial
support from the Academy of Finland grant Nos.\
136189, 140970 (PJK), 218159 and 141017 (MJM),
the University of Helsinki research project
`Active Suns', the Swedish Research Council grant
621-2007-4064, the European Research Council
under the AstroDyn Research Project 227952,
the EU COST Action MP0806, the European Research
Council under the Atmospheric Research Project
No.\ 227915, and a grant from the Government of
the Russian Federation under contract No.\
11.G34.31.0048 (NK,IR).
The authors thank NORDITA for hospitality during
their visits.


\end{document}